\documentclass[prb, aps, twocolumn, showpacs]{revtex4}
\usepackage{bm,amsmath}
\usepackage{graphics}
\usepackage{graphicx}
\usepackage{wrapfig,subfigure}
\usepackage{color,bigints}

\newcommand{\eg}{{\it e.g.~}}
\newcommand{\ie}{{\it i.e.~}}
\newcommand{\vb}{{$V_{\rm b}$~}}
\newcommand{\vg}{{$V_{\rm g}$~}}
\newcommand{\eF}{{\epsilon_{\rm F}}}
\newcommand\beq{\begin{equation}}
\newcommand\eeq{\end{equation}}

\begin{document}

\title{Conductance behavior in  nanowires with spin-orbit interaction -- A numerical study}
\author{Diego Rainis and Daniel Loss}

\affiliation{Department of Physics, University of Basel,
Klingelbergstrasse 82, CH-4056 Basel, Switzerland}


\begin{abstract}
We consider electronic transport through  semiconducting nanowires (W) with spin-orbit interaction (SOI), in a hybrid N-W-N setup where the wire is contacted by normal-metal leads (N). We investigate the conductance behavior of the system as a function of  gate and bias voltage,  magnetic field,  wire length, temperature, and disorder. The transport calculations are performed numerically and are based on standard recursive Green's function techniques.
In particular, we are interested in understanding if and how it is possible to deduce the strength of the SOI from the transport behavior.
This is a very relevant question since so far no clear experimental observation  in that direction has been produced.
We find that the smoothness of the electrostatic potential profile between the  contacts and the wire plays a crucial role, and we show  that in realistic regimes the N-W-N setup may mask the effects of SOI, and a trivial behavior with apparent vanishing SOI is observed.
We identify an optimal  parameter regime, with neither too smooth nor too abrupt potentials, where the signature of SOI is best visible, with and without Fabry-P\'erot oscillations, and is most resilient to disorder and temperature effects.
\end{abstract}
\maketitle

\section{Introduction}
The identification and the reliable assessment of the  spin-orbit-interaction (SOI) strength in semiconducting nanowires still represents an open problem. 
This very fact is even more surprising if one considers that SOI is a key ingredient in many contexts, \eg in spintronics~\cite{Awschalom_2002_book,Zutic_2004_RMP} 
and in spin-based quantum-computational schemes~\cite{Kloeffel_2013_Review}, and it is at the basis of one of the most studied proposals for Majorana fermions in condensed matter~\cite{AliceaReview,Sato,Lutchyn,bib:Oreg,bib:Exp1,bib:Exp2,bib:Exp3,Rokhinson_NPHYS_2012_Exp,Potter_PRB_2011,Klinovaja_Stano_PRL_2012,Rainis13PRB}, and more recently 
also for parafermions~\cite{Oreg_Sela_PF,Klinovaja_PF_PRB}, which provide promising platforms for topological quantum computing~\cite{Nayak_2008_RMP,AliceaReview}.

\begin{figure}[h!]
\includegraphics[width=1.0\linewidth] {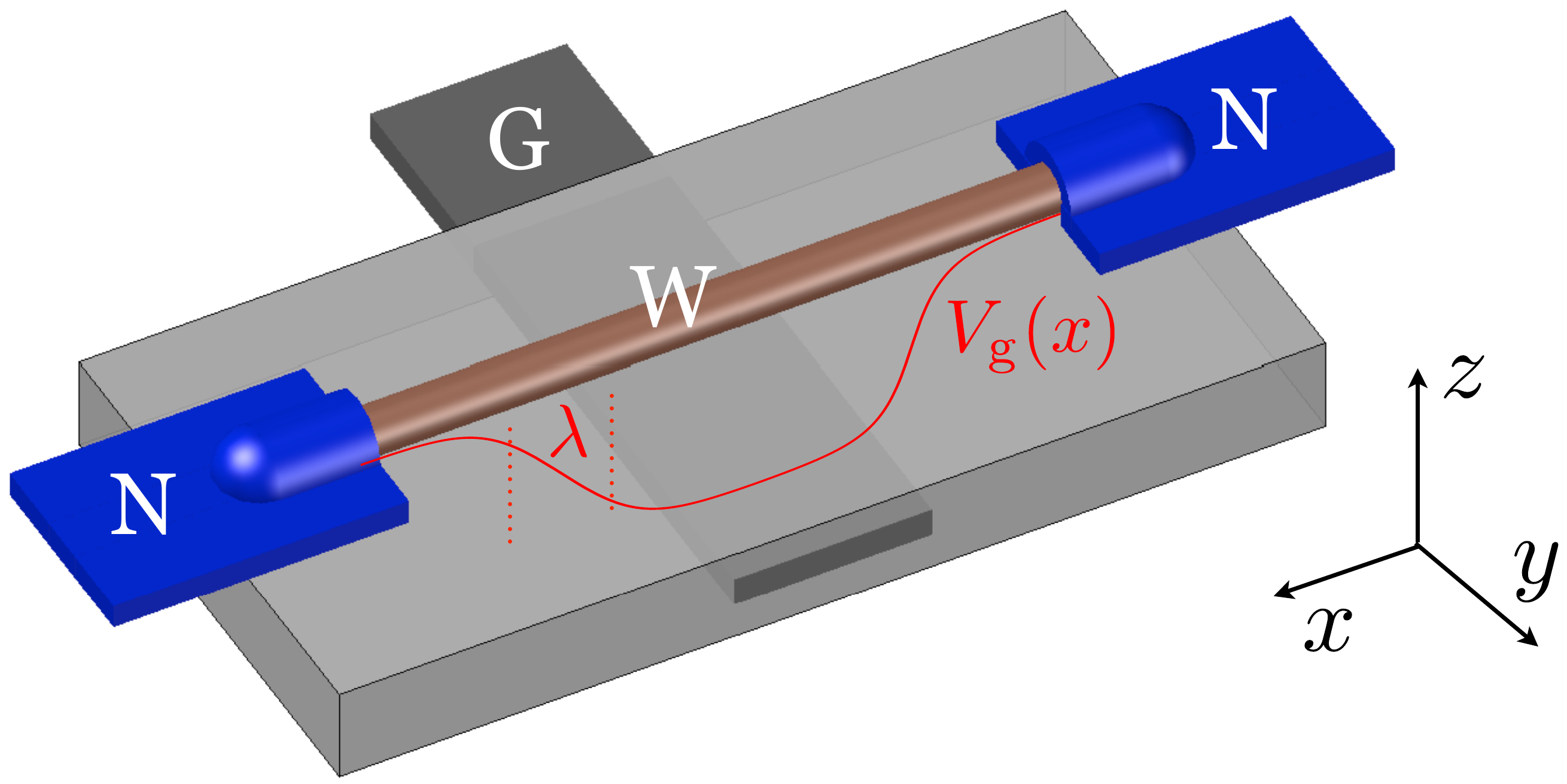}
\caption{(Color online) Typical N-W-N transport setup in which a nanowire (W)  is placed along the $x$-direction on a substrate lying in the $xy$-plane  and the two-terminal conductance $G$ of the wire is measured via normal electrodes (N, blue), while the chemical potential in the wire -- or better, in a {\it portion} of the wire -- is tuned via an underlying gate (G, dark gray). The actual electrostatic gate potential $V_{\rm g}(x)$ (red curve) is smooth and varies along $x$ on some typical length scale $\lambda$.
}
\label{fig_setup}
\end{figure}
\begin{figure*}
\includegraphics[width=0.75\linewidth] {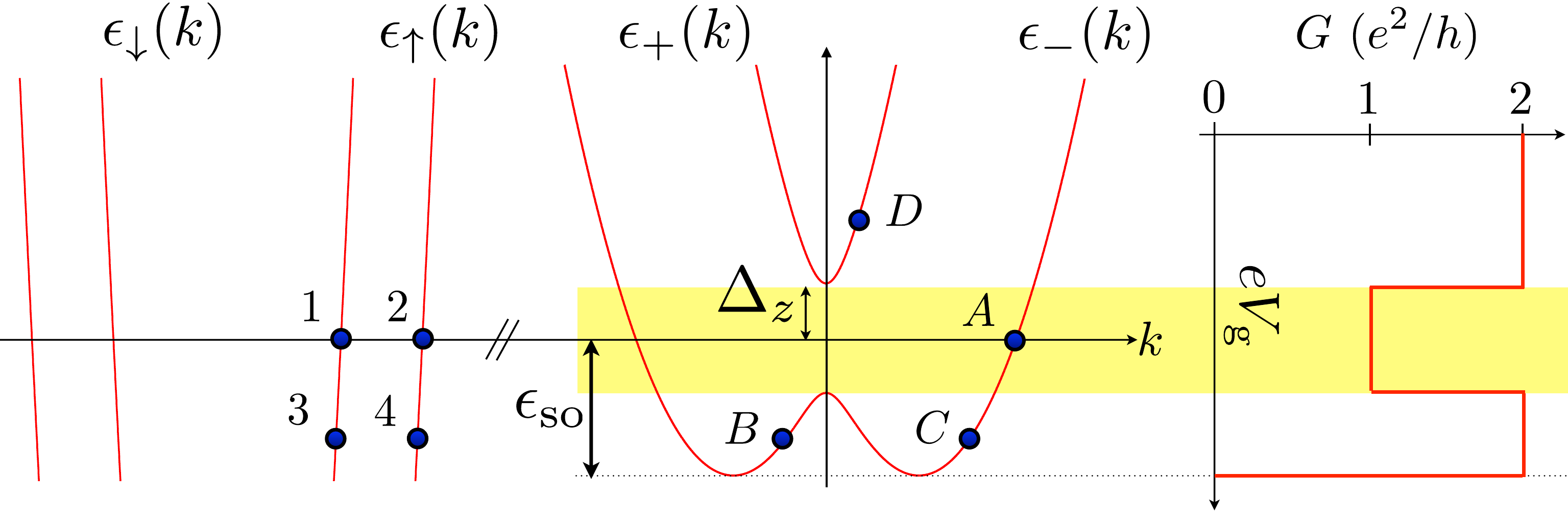}
\caption{(Color online) Expected SOI-induced re-entrant behavior in the W conductance. 
(Left) N dispersion, with two right movers per energy, labeled by $1,2$ and $3,4$ within and outside of the gap, respectively. (Middle) W-section dispersion, in the presence of Rashba SOI with energy $\epsilon_{\rm so}$ and $B$ field giving rise to the Zeeman gap $2\Delta_z$ at zero momentum $k$, highlighted in yellow (light grey), with a single right mover per energy ($A$).
Below the Zeeman gap there are two right movers per energy ($B,C$). (Right) Perfect conversion of $3,4$ into $B,C$ yields the ideal conductance behavior $G^{(0)}(V_{\rm g})$, where $G$ drops from $2e^2/h$ to $e^2/h$ within the Zeeman gap. 
}
\label{fig_SOI}
\end{figure*}
A typical measurement scheme involves the creation of quantum dots (QDs) in the nanowire (W)~\cite{FasthPRL2007,Nilsson_NL_2009_InSb_QD,Tarucha_NatNano_2011_InAs_QD,Leo_PRL_2012_InSb_QD,Leo_Nature_2013_NW_SOI_qubit_InAs,Leo_PRL_2013_Fast_Qubit_InSb,Leo_NatNano2013_NW_QD_Hole}, 
whereby the SOI strength is deduced from the observed  energy splittings (anticrossings) of the different spin states~\cite{FasthPRL2007,Stepanenko_2012_PRB}. 
Similarly, local and time-dependent Rashba SOI (induced by finger gates) can be accessed via transport in wires~\cite{Sherman_2013_PRB}.
This represents, however, only an indirect and not reliable measurement, especially because the SOI gets induced or modified by the QD geometry itself via the local electric fields of the confining gates. That is to say, in such experiments one measures the actual SOI for states in the dot, rather than the intrinsic SOI of the wire, which is the quantity one would like to know. Also, further complications in the spectrum of quantum dots arise when the SOI is unusually strong~\cite{Rashba_2012_PRB}.

This issue becomes even more relevant if one considers that most  wires are grown along the [111] direction, for which the Dresselhaus SOI induced by bulk-inversion-asymmetry~\cite{Dresselhaus_1955_PR} is expected to vanish~\cite{FasthPRL2007,Winkler_book}, and only the Rashba  SOI \cite{Rashba_1960_Sov} induced by structure-inversion-asymmetry should contribute -- if there are any SOI contributions at all. 
The latter is clearly strongly dependent on the particular setup, and is most likely different for a wire and a QD in the same wire.
Also, even if the wires possess finite Dresselhaus SOI, such as zincblende nanowires grown along the [110] direction \cite{FasthPRL2007,Winkler_book}, applying additional gates or metal contacts can generate Rashba SOI so that the total SOI can vary substantially along the wire.

A possible measurement scheme involving only the bare wires (without QDs) is based on the transport behavior of a SOI wire in the presence of a magnetic field $\bm B$ orthogonal to the SOI direction, where the conductance should show distinctive steps as different portions of the band dispersion are probed~\cite{StredaPRL2003,Pershin_PRB_2004_SOI_wires,Nagaev_2014_PRB}.
This idea is depicted in Fig.~\ref{fig_SOI}, where  the normal-metal-lead (N) dispersion (left panel), the wire dispersion (middle panel) and the corresponding ideal conductance $G^{(0)}$ (right panel) are shown.

The N sections of the model can play the role of the normal metal leads (with no SOI and lower $g$ factor) or simply of non-tuned sections of the wire ({\ie with a large local Fermi energy). The main results of the paper apply to both situations.
The N spin-split subbands are characterized by large velocities and two electronic states per propagation direction at each energy, labeled in Fig.~\ref{fig_SOI} by $1,2$  and $3,4$ for different energy regimes. 

In the nanowire, the interplay of SOI and $B$-field produces two spin subbands $\epsilon_\pm(k)$, with a characteristic  band folding at the lowest energies, where $\epsilon_-(k)$ exhibits two states per propagation direction at fixed energy (labeled by $B,C$ in Fig.~\ref{fig_SOI}).
Between such an energy range and $\epsilon_+(k)$ there is a partial gap of magnitude $2\Delta_z$ -- called Zeeman gap or helical gap -- where only one state per propagation direction is available (circle $A$), and the spin direction in the limit of strong SOI is,  to a first approximation, opposite for the two counterpropagating states (hence the term helical).
The energy separation between the band bottom and the midgap is given by the spin-orbit energy $\epsilon_{\rm so}=m^*\alpha^2/2\hbar^2$. 

The ideal conductance $G^{(0)}(V_{\rm g})$ (shown in the right panel of Fig.~\ref{fig_SOI}) shows the characteristic reentrant behavior, dropping from $2e^2/h$ to $e^2/h$ within the Zeeman gap. 
Here, $e$ is the electron charge and {$h=2\pi \hbar$} is Planck's constant.
An optimal situation is realized when the Zeeman gap and SOI energy are comparable, $\epsilon_{\rm so}\simeq 2\Delta_z$, so that the widths of the $2e^2/h$ step and of the $e^2/h$ plateau are both sizable and maximally visible (in contrast, for spin-filtering applications the limit $\Delta_z\ll \epsilon_{\rm so}$ is desirable).

Such ideal behavior is obtained either 
in a uniform {\it wire-only} calculation or in an ideal N-W-N setup where all wire states ($B,C$) are for some reason perfectly coupled to the lead states (3,4).
As the Landauer-B\"uttiker formalism highlights, $G^{(0)}$ in units of $e^2/h$ then just trivially counts the number of open channels $N_{\rm ch}^{\rm op}$, since the transmission probability of every channel is 
unity for open channels and zero for closed ones ({\it i.e.}, not crossing the chemical potential),
\beq
G^{(0)} = \sum_{n=1}^{N_{\rm ch}} T_n^{(0)} = N_{\rm ch}^{\rm op}\, .
\eeq
In such a case, $G^{(0)}$ would indeed provide a clean and direct method for characterizing SOI in nanowires, 
but so far the experimental results~\cite{Ilse_NL_2012_Conductance} have not shown the expected fingerprint of Fig.~\ref{fig_SOI}, 
or their  interpretation is still ambiguous and far from being conclusive~\cite{bib:Quay2010NatPhys}.
The same peculiar structure induced by SOI and Zeeman coupling has been shown to induce interesting spin filtering effects in hybrid junctions\cite{StredaPRL2003,Li_JAP_2014_spin_filtering}.

The main scope of this work is to focus on the mentioned hybrid N-W-N transport measurement setup, and to investigate theoretically possible mechanisms that might mask the expected conductance signature, and at the same time to propose an optimal configuration where the SOI-induced features should be maximally visible. 
As we will see,  a crucial role is played by the contacts, that is to say whether the electrostatic profile from the metallic leads to the semiconducting wire is slowly changing (adiabatic), abrupt, or intermediate. 
It is only in the latter case that transport can reveal the SOI optimally, while SOI can be completely invisible in the conductance as a function of gate voltage for the other cases. The crucial length scale separating the two  regimes is given by the ``critical Zeeman length'', $\lambda^\star={\hbar v_{\rm F}}/{\Delta_z}$ [see Eq.~(\ref{lambda}) below]. 
Finally, we show that the presence of Fabry-P\'erot oscillations in finite wires, which can emerge either as a function of gate voltage or magnetic field, can help in the detection of the SOI-induced features. 

This paper is organized as follows. In Sec.~\ref{Model} we introduce the model used to describe the transport properties of the one-dimensional wire. The model is  then evaluated numerically by standard recursive Green's function techniques.
In Sec.~\ref{Results} we present our results, with different sections referring to different setups or different physical regimes. Finally, in Sec.~\ref{Conclusions} we conclude by summarizing the results and commenting on the experimental situation.

\section{Model}
\label{Model}
The type of setups considered here is schematically shown in Fig.~\ref{fig_setup}. 
A semiconducting nanowire (W) with spin orbit interaction is contacted by two normal-metal leads (N), to which the bias voltage \vb can be applied. A wide gate below the nanowire with applied voltage \vg can be used to {\it locally} deplete it and to change its Fermi energy $\eF$, while additional narrow gates can be used to create sharp tunnel barriers at the ends of W.
The main results are captured by modeling the system through a  real-space tight-binding Hamiltonian in two dimensions,
\begin{align}
  \label{H}
  H = & \sum_{\bm i,\bm \delta} c^\dagger_{\bm i+\bm{ \delta},\alpha}
  \left[ -t \delta_{\alpha\beta}-i\bar\alpha_{\bm i}
  (\bm{\hat x}\cdot\bm{\delta})\sigma^y_{\alpha\beta} \right]c_{\bm{i},\beta}\nonumber\\
  +&\sum_{\bm i}c^\dagger_{\bm i,\alpha}\left[
  \epsilon_{\bm i}\delta_{\alpha\beta} - \frac{g_{\bm i}}{2}\mu_{\rm B}
  B_x\sigma^x_{\alpha\beta} \right]c_{\bm i,\beta} \;,
\end{align}
where we use standard second quantization notation for the fermionic creation ($c^\dagger_{\bm i,\alpha}$) and annihilation ($c_{\bm i,\alpha}$) operators,  and where
$t=\hbar^2/(2m^*a^2)$ is the hopping amplitude, chosen as energy unit, with $a$ being the lattice constant and $m^*$ the electron effective mass.
Further, $\bar\alpha$ is the spin-flip hopping amplitude, related to the physical SOI parameter by $\bar\alpha=\alpha/2a$ and to the SOI energy by $\epsilon_{\rm so}=\bar\alpha^2/t$ (zero in the leads and finite
and uniform in the wire). 
Throughout the paper we have set $\bar \alpha=0.01t$.
The sums run over all lattice sites $\bm i=(x,y)$ and  nearest neighbors $(\bm i+\bm \delta)$, and $\bm{\hat x}$ is a unit vector pointing along the wire axis. Implicit summation over repeated spin indices $\alpha,\beta=\uparrow, \downarrow$ is assumed. 

Further, the onsite energy $\epsilon_{\bm i}=V_{\rm g}({\bm i})+w({\bm i})$ accounts for the presence of the gate-induced electrostatic potential $V_{\rm g}$ (see Fig.~\ref{fig_setup}) and includes an on-site random potential $w$ (with rectangular probability distribution) that models Anderson disorder.
The gate potential $V_{\rm g}$, which plays the central role in this analysis, is instead modeled with a smooth profile along the wire that around each N-W contact or junction  located at $ x= x_0$  has the form (here specified for the left contact)
\beq
V_{\rm g}(x) = V_{\rm g}^{\rm N}+ \frac{V_{\rm g}^0- V_{\rm g}^{\rm N}}{2}\left[\tanh \left( \frac{ x-x_0 }{\lambda}  \right)+1\right] \;.
\label{gate_potential}
\eeq
This potential profile varies from the value $V_{\rm g}^{\rm N}$ in the leads (or far away from the N-W contacts) to the value $V_{\rm g}^0$ in the wire, and exhibits a linear behavior with slope $\Delta V_{\rm g} /\lambda$ around $x=x_0$ (we set $\Delta V_{\rm g}\equiv V_{\rm g}^0- V_{\rm g}^{\rm N})$. An example of such a profile is shown below in Fig.~\ref{fig_LZS} (for the right W/N contact).
The parameter $\lambda$  turns out to play a decisive role in determining the transport properties of the whole system.
The results are qualitatively independent on the specific value of $V_{\rm g}^{\rm N}$, but we will always choose large filling in the N sections, and to be specific we will  take $V_{\rm g}^0$ such that the chemical potential in the wire is tuned in the middle of the Zeeman gap.
 
Finally, the  magnetic field $\bm B$ points along the nanowire axis ($ \hat {\bm x}$) and induces a position-dependent Zeeman coupling $\Delta_{z}={g_{\bm i}\mu_{\rm B}}B/2$, where $g_{\bm i}$=2 in the metal leads (if present) and $g_{\bm i}=g^*$ in the wire ($\simeq$50 for bulk InSb, 15 for bulk InAs). No orbital effects are considered here.

We have numerically obtained the two-terminal conductance $G$=${\rm d}I/{\rm d}$\vb, with $I$ the charge current and \vb the applied bias voltage, of the above described hybrid structure by employing the standard scattering theory~\cite{bib:LambertRaimondi1998}, with the help of the recursive Green's function techniques~\cite{bib:RecursiveGFMacKinnon}, similarly to what has been done in our previous work~\cite{Rainis13PRB, Rainis_2014_PRL_FF,Saha_Rainis_2014_PRB}. 
We calculate $G$ as a function of the gate potential \vg (at fixed zero bias \vb$=0$), but the results  should be essentially identical to the conductance measured as a function of bias voltage \vb (measured at fixed zero gate potential \vg$=0$).

\begin{figure}[h!]  
\includegraphics[width=1.0\linewidth] {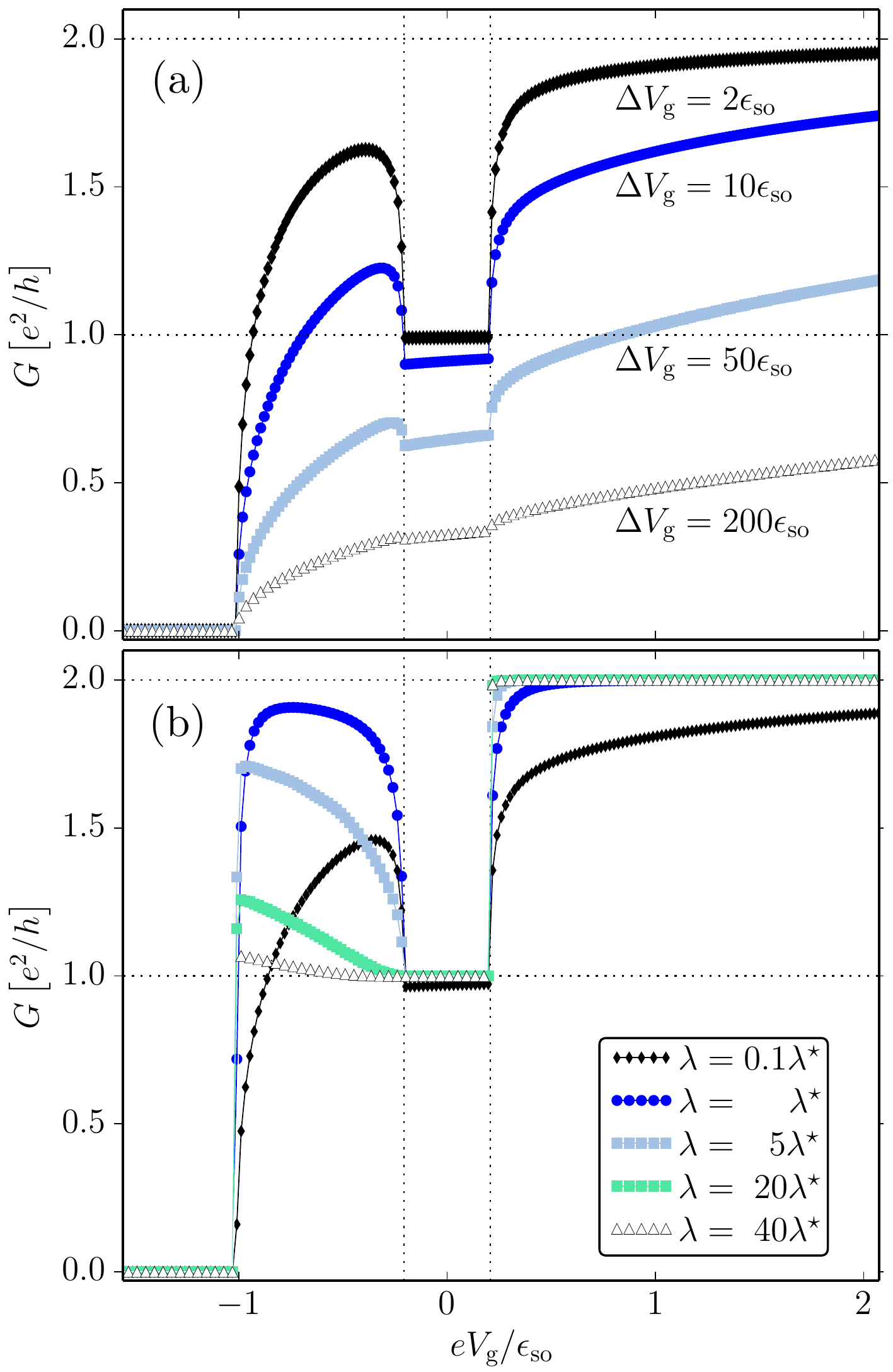}
\caption{Zero-temperature conductance behavior in the simple N-W measurement setup (semi-infinite wire with only one contact) as a function of gate voltage \vg, with profile given in Eq.~(\ref{gate_potential})
and for typical parameter values $\epsilon_{\rm so}=50~\mu$eV and $\Delta_z=10~\mu$eV [see text after Eq.~(\ref{lambda})].
(a) Abrupt case $\lambda=0$. Different curves correspond to different total variations of \vg across the junction.
In the case of large  gate voltage variations ($\Delta$\vg =200$\epsilon_{\rm so}\simeq 10$meV), the conductance $G$ is largely reduced and almost no signature of SOI appears compared to the ideal case shown
in Fig.~\ref{fig_SOI}.
By decreasing the induced Fermi energy variations one gets a $G(V_{\rm g})$ curve that exhibits the desired reentrant behavior. The vertical dotted lines indicate the extension of the Zeeman gap.
(b) Dependence of $G$ on the smoothness $\lambda$ of the gate potential profile, at fixed $\Delta V_{\rm g}=10\epsilon_{\rm so}\simeq0.5$ meV. At some optimal value $\lambda\simeq\lambda^\star$ the conductance shows the characteristic SOI-induced step at $2e^2/h$, but for increasing values of $\lambda$ the step is flattened {down} to a single $e^2/h$ plateau, like in the case of SOI=0 (and finite Zeeman gap).
}
\label{fig_Lambda_semiinf}
\end{figure}

\section{Results}
\label{Results}
We show in this section how the smoothness $\lambda$ of the electrostatic gate potential profiles \vg$(x)$ influences the conductance $G$ behavior of the setup.
We first consider a single N-W contact, that is, a semi-infinite wire, in order to avoid finite-size effects and focus on the intrinsic properties of the junction.
Subsequently, we consider a realistic, finite wire in a N-W-N setup and investigate in some detail the ensuing Fabry-P\'erot conductance oscillations.

\subsection{N-W -- Abrupt Contacts}
We first start from the extreme case of abrupt contacts [Fig.~\ref{fig_Lambda_semiinf}(a)], where all the position-dependent quantities change over the length scale of the lattice constant $a$. 
It may be considered an unrealistic situation, but it can prove instructive in this conceptual investigation, also because it allows for an analytical solution
(see similar results in Ref.~\onlinecite{StredaPRL2003}).
The conductance $G(V_{\rm g})$ in this case is considerably reduced compared with the quantized value, and the reduction is stronger for the ``internal'' states ($B$ and $D$ in Fig.~\ref{fig_SOI}), as one can deduce from the small conductance jumps at the gap edges in the light blue and white curves of Fig.~\ref{fig_Lambda_semiinf}(a).
The reduction changes for different total Fermi energy variations $\Delta V_{\rm g}$, giving rise to well-defined SOI steps only for $\Delta V_{\rm g}$ comparable to $\epsilon_{\rm so}$, see dark blue and black curves of Fig.~\ref{fig_Lambda_semiinf}(a).
The numerical results are in full agreement with the analytical ones obtained by a  wave functions matching method.

%%%%%%%%%%%%%%%%%%%%
\subsection{N-W -- Smooth Contacts and Adiabatic Limit}
Considering instead more realistic smooth N-W contacts  and varying the steepness $\lambda$ of the gate potential \vg, given in Eq.~(\ref{gate_potential}), one observes an unexpected behavior, which represents the main focus of this work.
As seen above,  for abrupt or steep gate potentials, the conductance $G$ shows a non-ideal behavior [black diamonds in Fig.~\ref{fig_Lambda_semiinf}(b)], featuring more or less visible gap-related signatures depending on the value of $\Delta V_{\rm g}$.
For smoother gate potentials, characterized by intermediate values of $\lambda$ [dark blue circles in Fig.~\ref{fig_Lambda_semiinf}(b)], the curve $G(V_{\rm g})$ almost reaches the ideal shape (2--1--2)$e^2/h$,  fully revealing the Zeeman gap and the SOI-induced $2e^2/h$ step at the lowest energies [dark blue circles and, partially, light blue squares in Fig.~\ref{fig_Lambda_semiinf}(b)].
\begin{figure}[h!]
\includegraphics[width=1.0\linewidth] {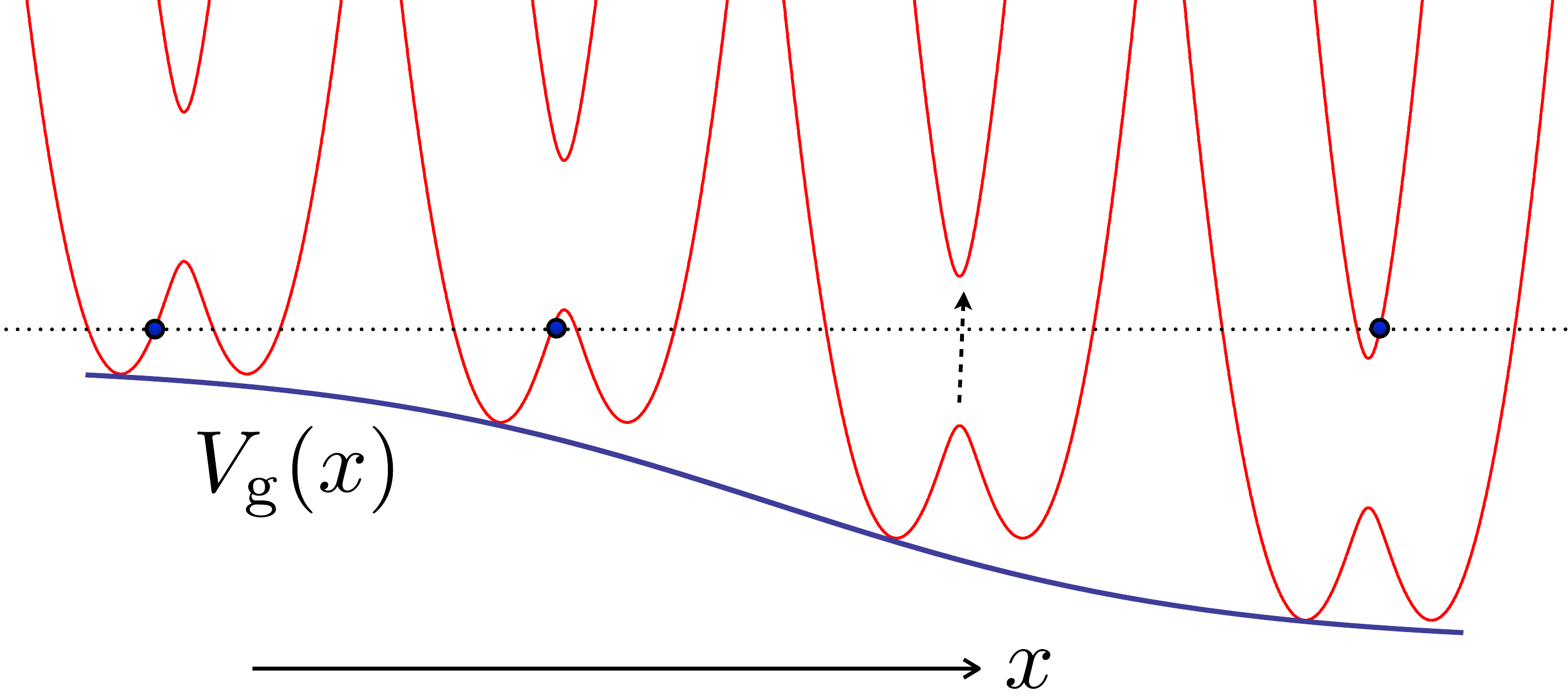}
\caption{(Color online) Real-space Landau-Zener--St\"uckelberger-like mechanism that determines the visibility of the SOI step in the adiabatic limit. 
As the gate potential $V_{\rm g}(x)$ varies along the wire (blue solid curve), the local dispersion relation is vertically shifted and different states are involved at the given energy (dotted line). There exists a finite region where the chemical potential lies within the Zeeman gap (third picture from the left), and the tunneling through this forbidden region gives rise to  Landau-Zener-St\"uckelberger physics.
}
\label{fig_LZS}
\end{figure}

Finally, let us consider large values of $\lambda$, such that all parameters change very slowly in space ($x$-direction), over distances much larger than any length scale present in the problem.
This situation corresponds to the {\it adiabatic} limit, where every incoming scattering mode -- when possible -- smoothly evolves into a propagating mode in the wire. 
For instance, for a quantum-point-contact geometry it is known that in such a limit  every incoming channel is either fully transmitted or fully reflected, depending on its energy~\cite{Buttiker_PRB_1990_Saddle}.
The result would be identical for a nanowire in the absence of a magnetic field, where each electronic state can smoothly evolve as $V_{\rm g}$ is varied.
In our case with SOI, the situation is complicated by the fact that the propagating states on either side of the junction are instead  not smoothly connected, due to the presence of the Zeeman gap.
As schematically shown in Fig.~\ref{fig_LZS}, states at the chemical potential that in a certain region belong to the $\epsilon_-$ band need to end up into the $\epsilon_+$ band in a different spatial region, where the gate potential \vg is weaker. This implies that there exists an intermediate region where the chemical potential lies within the Zeeman gap, and the ``internal'' electron has to tunnel through such a region in order to contribute to the conductance.
This tunneling process gives rise to a sort of spatial Landau-Zener-St\"uckelberger  mechanism, generated by a spatial rather than a temporal variation of a system parameter (here Fermi energy).
The critical length $\lambda^\star$ that enters the adiabaticity condition can be identified as
\beq
\lambda^\star=\frac{\hbar v_{\rm F}}{\Delta_z}\;,
\label{lambda}
\eeq
where $v_{\rm F}$ in our case is the Fermi velocity in the zero-field case ($v_{\rm F}=\alpha/\hbar$), and it serves just to provide an estimate for $\lambda^\star$ that is valid for most energies within the relevant SOI-induced step. 
The case $\lambda\simeq\lambda^\star$ represents an optimal condition, for which the SOI-related $2e^2/h$ plateau is maximally visible.
For the values considered here, $\epsilon_{\rm so}=50~\mu$eV in InSb and $B=10$ mT, approximately corresponding to $\Delta_z=10~\mu$eV, we find $\lambda^\star\simeq1~\mu$m.
For $\Delta_z=\epsilon_{\rm so}$ one finds that $\lambda^\star=2\ell_{\rm so}\simeq400$ nm here. 
A transport configuration is expected to be adiabatic when the characteristic  smoothness of \vg$(x)$ [see  Eq.~(\ref{gate_potential})] satisfies $\lambda\gg\lambda^\star$, and to be non-adiabatic in the opposite limit $\lambda\ll\lambda^\star$.

The numerical results indeed show that for very slowly changing gate potentials [green circles and white triangles in Fig.~\ref{fig_Lambda_semiinf}(b)] the contribution from the internal state becomes less and less relevant, and the conductance $G$ eventually exhibits a monotonic behavior as a function of \vg with a single, joint plateau at $e^2/h$ at the lowest energies. 
In such a case one does {\it not} have the possibility to read off the value of the SOI from the conductance behavior, 
and the approach of using transport as a way to determine the SOI strength no longer works.

We remark that $\lambda^\star$ is, in principle, an energy-dependent quantity, since the electronic velocity is energy-dependent, and indeed in all results we observe different conductance reductions for different energies (\ie different $V_{\rm g}$), see Fig.~\ref{fig_Lambda_semiinf}(a).

Finally we note that  we have assumed that the SOI always changes abruptly from the wire to the normal leads. However, we also checked smooth transitions for the SOI but found no visible effect on the conductance in the regime of interest here. This is to be expected since where $\alpha$ would change smoothly the Fermi energy is already so large that SOI effects are negligible. 

\begin{figure}[h!]  
\includegraphics[width=0.90\linewidth] {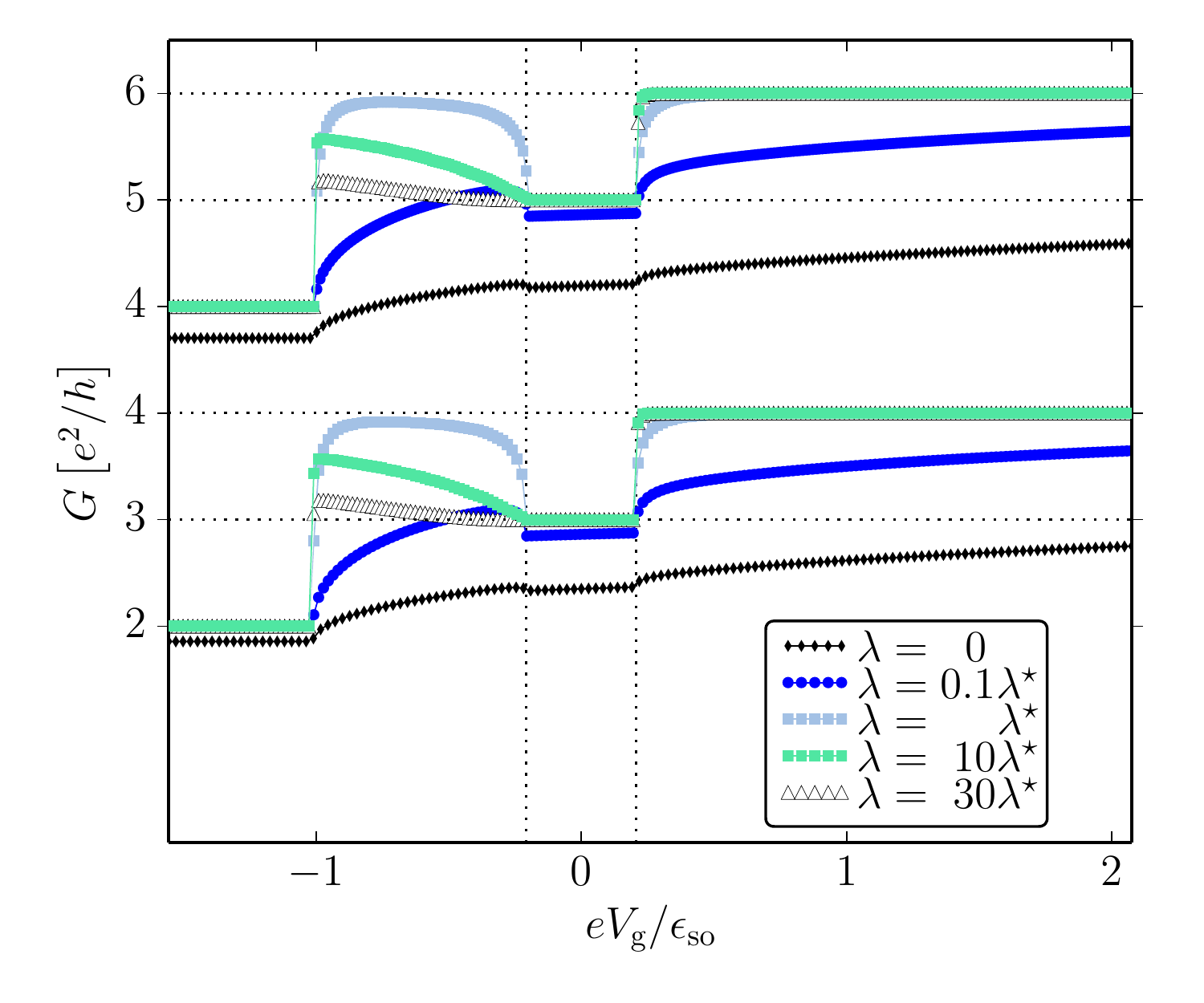}
\caption{Zero-temperature conductance behavior ($G$ in units of $e^2/h$) in the case of two (lower curves) and three (upper curves, shifted by $e^2/h$) occupied subbands. 
$G$ shows the desired reentrant behavior in an intermediate regime $\lambda\simeq \lambda^\star$, while for either too steep (black diamonds and blue circles) or too smooth (green squares and white triangles) gate potential profiles the SOI-related step becomes less and less visible. 
}
\label{fig_Multichannel}
\end{figure}
%

%%%%%%%%%%%%%%%%%
\subsection{N-W -- Multi-Channel Case}
We have checked that the observed reduction of conductance is not specific to the one-channel limit, and that in the multi-channel case each individual subband provides a similar contribution, with conductance reductions depending on the corresponding Fermi energies. For a visualization of the results, see Fig.~\ref{fig_Multichannel}, where we report the conductance curves in the case of two and three occupied subbands.
As observed in the single channel case, $G$ has the ideal behavior only in some intermediate regime $\lambda\simeq\lambda^\star\simeq0.5-2~\mu$m, while for abrupt gate potential profiles \vg$(x)$ it shows an overall reduction (black diamonds and dark blue circles), and for too smooth gate potentials it exhibits a reduced contribution from the second, internal channel of the top-most band (green squares and white triangles). At the same time the lower bands, with larger Fermi energies,  almost always contribute with a full quantum of conductance per channel, apart from the abrupt case where the matching is not perfect (black diamonds).

%%%%%%%%%%%%%%%%%%%%%%%%
\subsection{N-W-N -- Fabry-P\'erot Resonances}
\begin{figure}[h!]  
\includegraphics[width=0.90\linewidth] {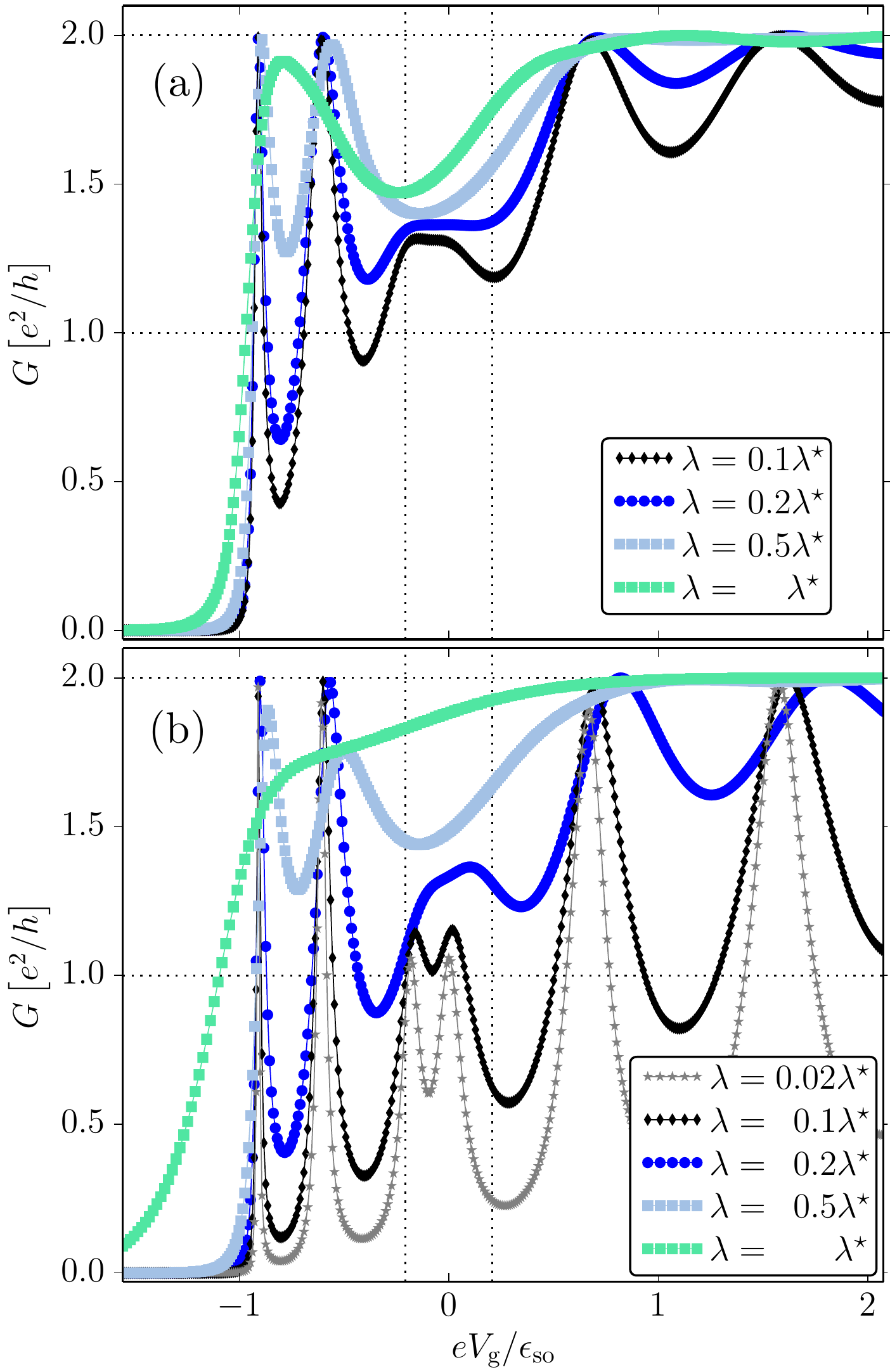}
\caption{Zero-temperature conductance behavior in a N-W-N setup with wire length $L=2~\mu$m. 
(a) 
Dependence of $G(V_{\rm g})$ on the smoothness $\lambda$, at fixed $\Delta V_{\rm g}=5\epsilon_{\rm so}$.
(b)
Same as in (a), but with different $\Delta V_{\rm g}=50\epsilon_{\rm so}\sim$ meV's.
}
\label{fig_FP_100}
\end{figure}
\begin{figure}[h!]  
\includegraphics[width=0.90\linewidth] {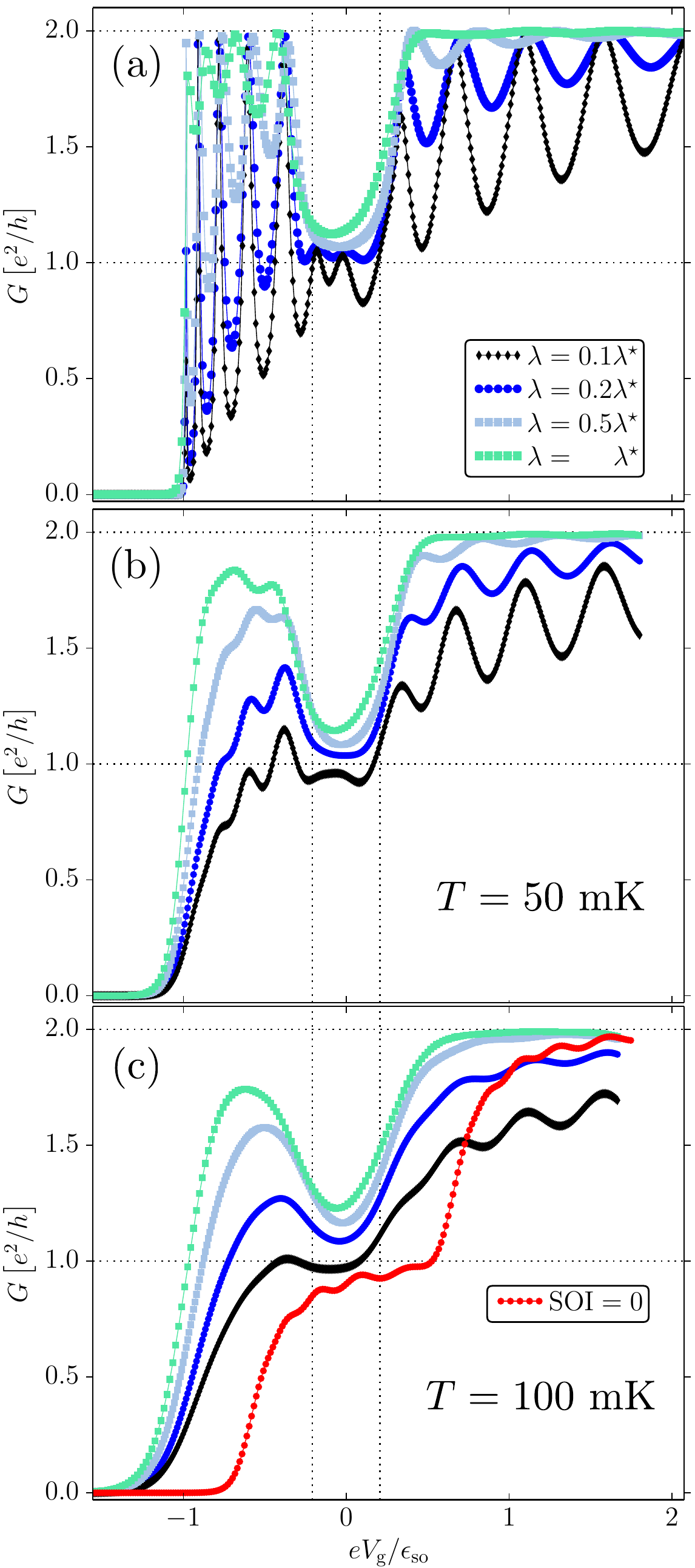}
\caption{Same as in Fig.~\ref{fig_FP_100}, but for a twice as long wire, $L=4~\mu$m, and intermediate $V_{\rm g}=10\epsilon_{\rm so}$. 
(a) 
$T=0$ conductance behavior, with Fabry-P\'erot resonances exhibiting halved energy spacing in comparison with those of Fig.~\ref{fig_FP_100}, as expected.
(b)
$T=50$ mK. The oscillations show reduced amplitude, and for the lowest values of $\lambda$ we get $G\simeq e^2/h$ within the SOI step.
(c)
$T=100$ mK. The lowest FP oscillations have disappeared and for $\lambda\ll\lambda^\star$ (in particular, the black curve, referring to $\lambda\simeq100$ nm) give rise to a single, almost flat $e^2/h$ plateau, masking any signature of SOI . Indeed the SOI=0 curve, plotted in red circles, shows a very similar behavior.
}
\label{fig_FP_200}
\end{figure}

If we now consider the realistic finite-length case of a N-W-N setup, Fabry-P\'erot (FP) oscillations are expected to appear in the conductance curve as a function of gate voltage.
This is indeed observed in the simulations in Fig.~\ref{fig_FP_100}, involving  a 2 $\mu$m-long wire, where we show that clear resonances are superimposed on the background contribution that retraces the semi-infinite result of Fig.~\ref{fig_Lambda_semiinf}.
The same main trend is observed when $\lambda$ is varied.
For the steepest potential profiles (black diamonds and dark blue circles), corresponding to $\lambda\ll\lambda^\star$, the conductance $G(V_{\rm g})$ is reduced compared to its quantized value (see the minima of $G$), and the reduction is more evident for larger $V_{\rm g}$ variations, compare Figs.~\ref{fig_FP_100}(a) and \ref{fig_FP_100}(b).

We note that, due to the smoothness of the potential $V_{\rm g}(x)$, the region where the gating is effective and produces a uniform potential is shorter than the nominal wire length and is approximately given by $L-2\lambda$.
One consequence is that the confining length becomes energy-dependent (\ie gate-dependent), and thus the energy spacing and the width of the FP resonances are nonuniform.
For $\lambda$ comparable to $L$ the FP oscillations have almost completely disappeared, see Fig.~\ref{fig_FP_100}.
Also, the effective one-channel region becomes so short that tunneling through it becomes appreciable and $G$ exceeds the $e^2/h$ value within the Zeeman gap, see Fig.~\ref{fig_FP_100}(a) and the $\lambda>0.1\lambda^\star$ curves in Fig.~\ref{fig_FP_100}(b).
This is another mechanism by which the SOI-induced reentrant behavior of the conductance is lost and SOI strength cannot be deduced from the conductance measurements.

Repeating the same calculation for a wire with twice the length, $L=4~\mu$m, we find denser Fabry-P\'erot resonances, as expected.
As one can observe in the zero-temperature curves of Fig.~\ref{fig_FP_200}(a),
another consequence of the increased length is that $G$ remains bounded by $e^2/h$ within the Zeeman gap and a reentrant behavior can in principle be obtained.
However, when considering the realistic finite temperature scenario, we find that the SOI-induced feature can be lost.
In particular, for $T=50$ mK the Fabry-P\'erot oscillations still survive but for $\lambda\ll\lambda^\star$ the average conductance is $\simeq e^2/h$ even in the SOI-induced two-channel regime, as shown in Fig.~\ref{fig_FP_200}(b).
For $T=100$ mK the FP oscillations have disappeared completely at the lowest energies and the resulting conductance curve for $\lambda\ll\lambda^\star$ exhibits a single, almost-flat $e^2/h$ plateau that resembles the zero-SOI result, see panel (c).
{To better highlight how difficult it could be to identify the presence of SOI from a conductance curve, we plotted in the same panel (red circles) also the conductance curve for an identical wire, but with SOI=0 (and finite Zeeman gap).
For this plot we actually chose a slightly larger Zeeman splitting and a smoother gating profile, in order to reproduce the width of the $e^2/h$ plateau of the black curve. However, since both the actual $g$-factor and the electrostatic profile are unknown in the experiment, it is in fact meaningful and necessary to investigate different setups that could potentially yield the same results.}

The interplay just described between finite-size Fabry-P\'erot oscillations and temperature smearing represents a third mechanism by which the SOI-induced conductance step may become effectively invisible in a realistic transport measurement.

%%%%%%%%%%%%%%%%%%%%%%%%
\subsection{{N-W-N -- Inclusion of Disorder}}
\label{disorder}

When one investigates the impact of realistic amounts of Anderson disorder (modeling random neutral impurities), one finds that the above described conductance pattern that allows the identification and the measurement of SOI strength is rather fragile, and it is preserved only in almost-ballistic wires. 
We considered the optimal regime $\lambda=\lambda^\star$ and analyzed the conductance behavior of wires with $L=4~\mu$m, in order to avoid the short-wire effects present when $\lambda\sim L$ (as described in the previous section).
Relying on the most recent estimates of disorder in InSb nanowires~\cite{Plissard_2012_NL_InSb}, we assume $\ell_{\rm mfp}\simeq300$ nm and we consider different disorder configurations, without averaging (quenched disorder).
In Fig.~\ref{fig_FP_Disorder}(a) we consider such regime, with a temperature of $T=50$ mK. 
We plot the conductance curve of three different representative disorder configurations, and 
in the same panel we also report the clean-wire result of Fig.~\ref{fig_FP_200}(b), with the same color code (green circles).
It is apparent that for all three configurations the SOI-induced reentrant behavior is absent or at least not at all recognizable, due to the appearance of disorder-induced resonances and to an overall sensible reduction of $G$. 
The severe deterioration of the sought-for feature in the conductance curves is mainly due to 
the different way disorder impacts on the two conducting channels.
While the high-velocity external states induce a low density of states (DOS), the internal states under examination here exhibit low-velocities and large DOS, leading to a larger impact of disorder. This is indeed confirmed by the fact that within the Zeeman gap disorder induces smaller reductions and more controlled sample-to-sample fluctuations.
We then consider milder disorder, corresponding to $\ell_{\rm mfp}\simeq 2-3~\mu$m, in order to explore the quasi-ballistic regime. We find that for some disorder configurations $G(V_{\rm g})$ still bears some resemblance to the clean-case curve, and SOI could be detectable. This is seen, for example, in the blue curve in Fig.~\ref{fig_FP_Disorder}(b).
Already at $T=100$ mK, however, the feature is partially washed out and could  easily be confused with Fabry-P\'erot oscillations, see Fig.~\ref{fig_FP_Disorder}(c).

\begin{figure}[h!]  
\includegraphics[width=0.90\linewidth] {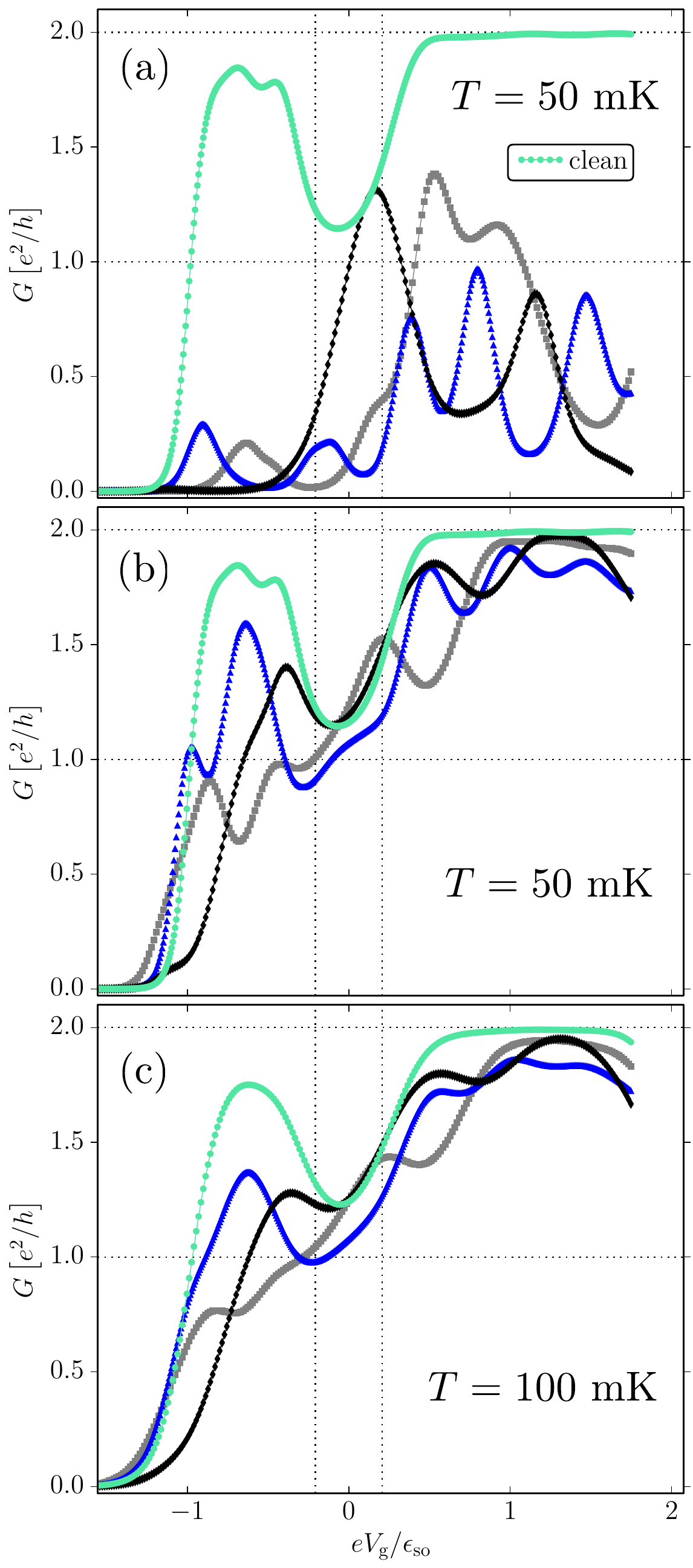}
\caption{Conductance for a finite wire of length $L=4~\mu$m in the presence of disorder, with optimal condition $\lambda=\lambda^\star$. The clean-case result is shown as green circles,  all other curves refer to different disorder configurations.
(a) Realistic mean free path values ($\ell_{\rm mfp}\simeq300$ nm) destroy the SOI-induced features, and no indication of SOI can be read off.
(b) With milder disorder ($L/\ell_{\rm mfp}\simeq1-2$) and $T=50$ mK it is still possible to observe the SOI step, depending on the disorder configuration.
(c) For $T=100$ mK the SOI peak already gets smeared out and is no longer uniquely identifiable anymore.
}
\label{fig_FP_Disorder}
\end{figure}

\subsection{Fabry-P\'erot oscillations as a function of $B$-field}
\label{FP_Bfield}
One can also consider an experimental situation in which the bias potential and the gate potential (wire doping level) are kept fixed, and the conductance is measured while the magnetic field amplitude is varied.
When  \vg is such that the chemical potential lies within the Zeeman gap -- or, equivalently, $\Delta_{z}>\epsilon_{\rm F}$ -- only the external, large-$k_{\rm F}$ states are involved in transport and regular single-period oscillations show up in $G(B)$, as shown in the right part of the plots of Fig.~\ref{fig_FP_B}.
In contrast, if the chemical potential lies outside the Zeeman gap (\ie $\Delta_z<\epsilon_{\rm F}$), two states per propagation direction are involved in transport and more complicated patterns are observed, due to the different energy evolution of the two subbands as $B$ is varied (left portion of the curves). 
This bi-modal behavior, absent in the usual $G(V_{\rm g})$ measurement, could be exploited to indirectly determine the SOI energy $\epsilon_{\rm so}$, given that, even in the adiabatic limit where the signal from the internal channel is weak, one can 
distinguish the two regimes and locate the $\Delta_z=\epsilon_{\rm F}$ point  by means of Fourier analysis.
The different plots of Fig.~\ref{fig_FP_B} correspond to different Fermi energies and different lengths (see the figure caption).

\begin{figure}[h!]  
\includegraphics[width=1.0\linewidth] {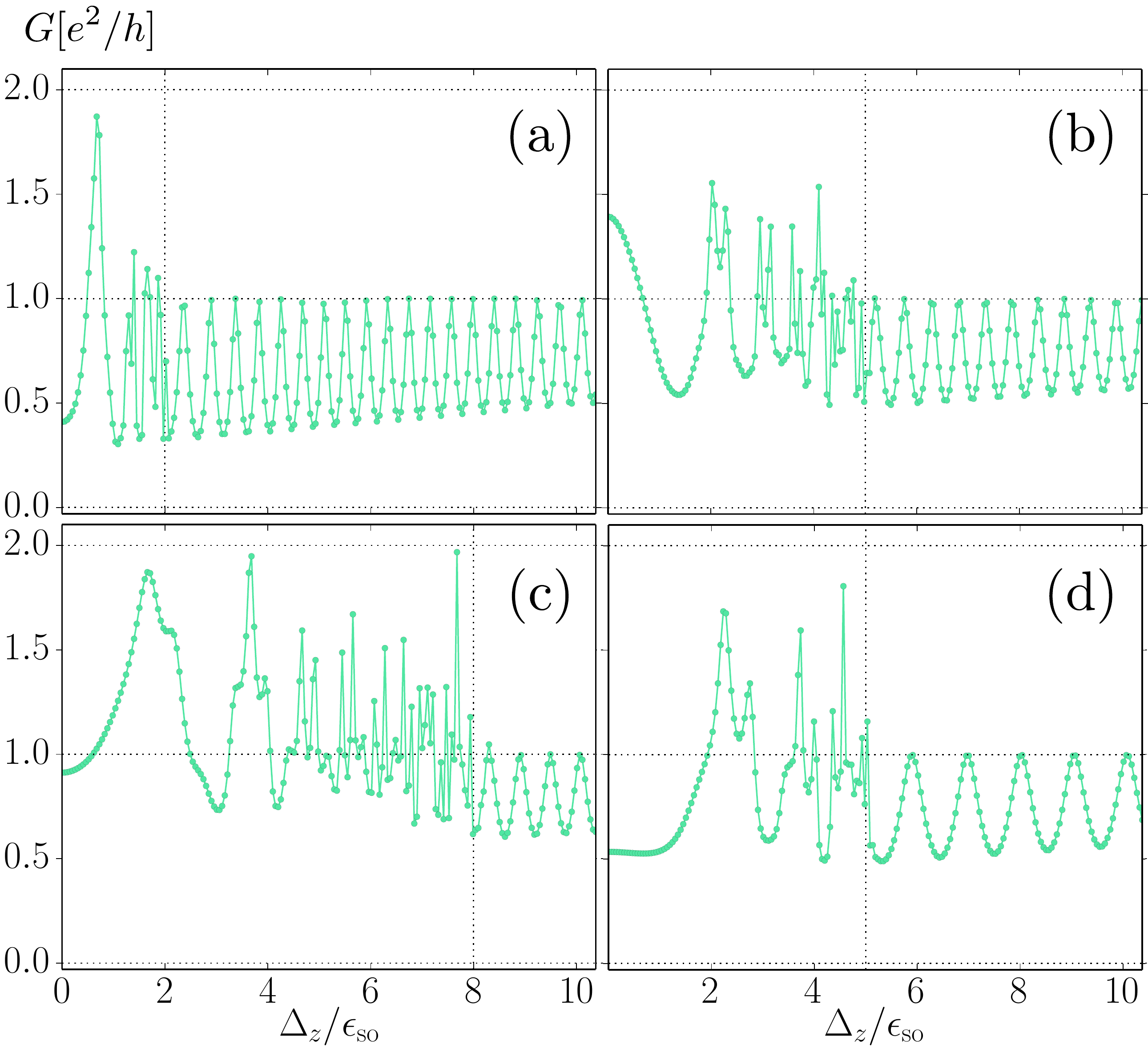}
\caption{Zero-temperature Fabry-P\'erot resonances observed in $G(B)$. When  $\Delta_{z}>\epsilon_{\rm F}$ only one state is involved in transport and regular single-period oscillations show up in $G(B)$, see the portion of the curves on the right of the vertical dotted line identifying $\Delta_z=\epsilon_{\rm F}$.
When the chemical potential lies outside the Zeeman gap two states are involved and more complicated oscillations are observed (left part of the curves).
The plots refer to the case $L=10~\mu$m. (a) $\epsilon_{\rm F}=2\epsilon_{\rm so}$,
(b) $\epsilon_{\rm F}=5\epsilon_{\rm so}$, (c) $\epsilon_{\rm F}=8\epsilon_{\rm so}$,
 and (d) $\epsilon_{\rm F}=5\epsilon_{\rm so}$ and $L=5~\mu$m. Note that this curve presents half-period oscillations, as expected [compare to panel (b)]. For all these plots we kept $\lambda=0.1\lambda^\star$.
}
\label{fig_FP_B}
\end{figure}

Upon inclusion of disorder, the (non-averaged) conductance behavior $G(B)$ may significantly change and may no longer show the bi-modal behavior.
In Fig.~\ref{fig_FP_B_dis} we show the results corresponding to two different disorder strengths, in a wire with $L=4~\mu$m and finite temperature $T=50$ mK.
In general, we observe that the single-periodic oscillations at high Zeeman fields are still well defined, while at low fields there are some larger modifications. 
This is in agreement with the different behavior of internal and external states  in the presence of disorder, as  discussed above.
In Fig.~\ref{fig_FP_B_dis}(a) realistic disorder corresponding to a mean free path $\ell_{\rm mfp}\simeq300$ nm is considered. 
Various curves corresponding to some representative disorder configurations are shown together with the clean case result (green curve). No major variations in $G(B)$ are observed.
In Fig.~\ref{fig_FP_B_dis}(b) we have considered strong disorder corresponding to $\ell_{\rm mfp}\simeq150$ nm. 
Even if  the curves related to different disorder configurations now deviate sensibly from one another, 
the two regimes are still clearly discernible and allow for a simple identification of the point $\Delta_z=\epsilon_{\rm F}$.

\begin{figure}[!] 
\includegraphics[width=0.9\linewidth] {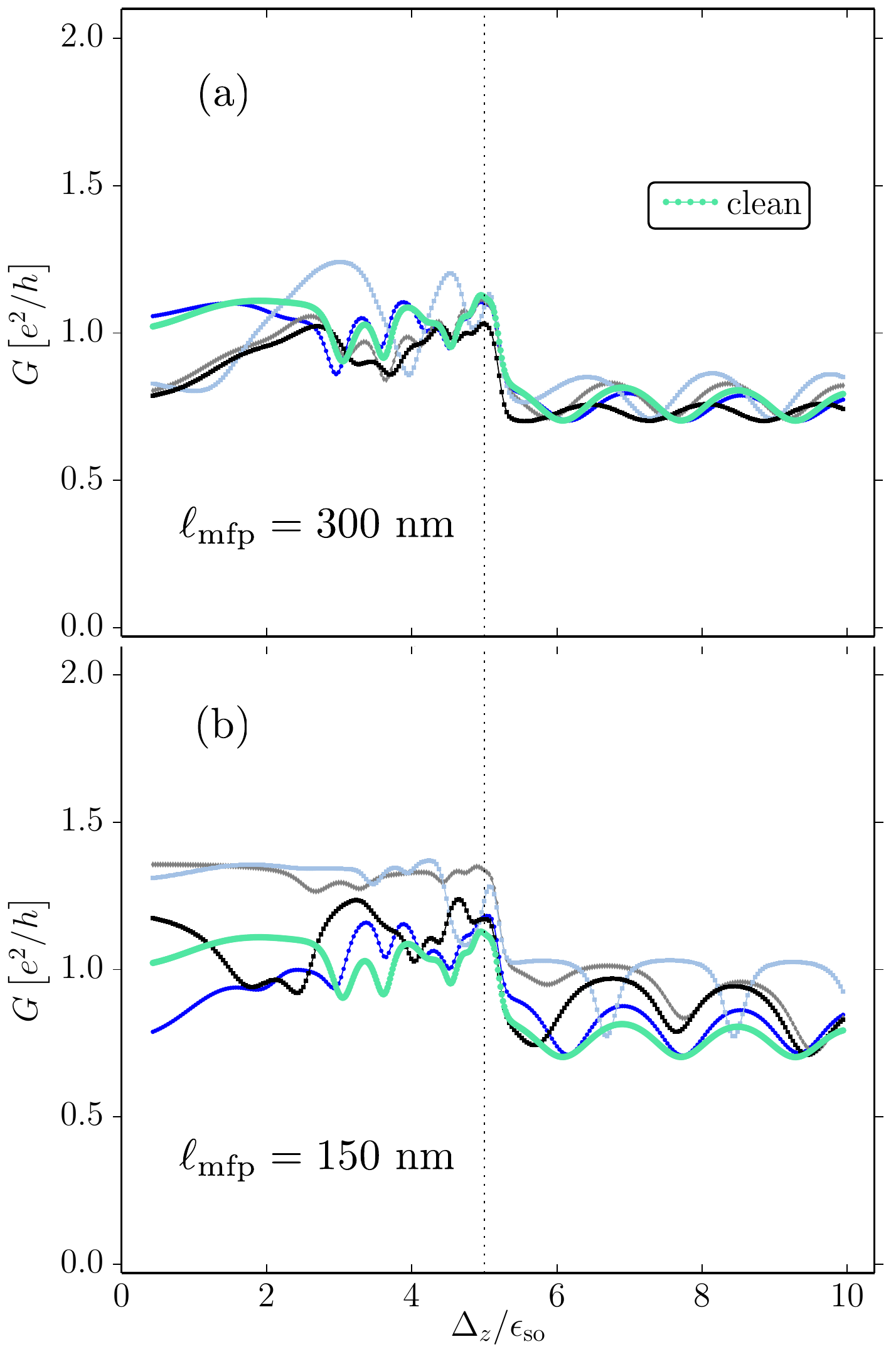}
\caption{Fabry-P\'erot resonances observed in $G(B)$ in the presence of disorder in a 4 $\mu$m-long wire, for $\epsilon_{\rm F}=5\epsilon_{\rm so}$, $\lambda=0.1\lambda^\star$ and  finite temperature $T=50$ mK. 
(a) Realistic disorder corresponding to a mean free path $\ell_{\rm mfp}\simeq300$ nm. Different curves refer to different disorder configurations. The green curve corresponds to the clean case result. $G(B)$ for different configurations shows few variations, especially in the single-channel regime $\Delta_z>\epsilon_{\rm F}$.
(b) Strong disorder $\ell_{\rm mfp}\simeq150$ nm. The difference between the two regimes is still  clearly discernible for each configuration, even if the different curves substantially  deviate from each other.
}
\label{fig_FP_B_dis}
\end{figure}

\section{Conclusions} 
\label{Conclusions} 
We have considered differential conductance measurements in hybrid N-W and N-W-N setups, where the electrostatic gate potential has a smooth profile, due to the presence of gating electrodes.
We have shown how the expected SOI-induced reentrant behavior in the conductance can be absent due to several mechanisms, which are all related to the smoothly varying gate potential $V_{\rm g}(x)$. We identified a  length scale $\lambda^\star=\hbar v_F/\Delta_z$ and showed that it allows one to separate adiabatic behavior from non-adiabatic behavior of the gate potential.
The masking effect is present  for both too small and for too large values of the potential smoothness $\lambda$ (compared to $\lambda^\star$), even though it is due to different mechanisms, while for an optimal value $\lambda\simeq\lambda^\star$ the SOI-induced step is still almost fully visible, even after inclusion of mild disorder and finite temperature. 
Our analysis shows that transport measurements must be carefully set up and analyzed in order to identify the SOI unambiguously. Alternative measurement methods such as tunneling and optical spectroscopy or indirect non-invasive measurements via nanoscale NMR of nuclear spins~\cite{Poggio_2013_Nuclear},
which  reveal the presence of SOI in the nuclear spin relaxation \cite{Zyuzin_Meng_PRB_2014}, might also be considered.

 \section*{Acknowledgments}
We acknowledge insightful discussions with Fabio Taddei and support by the Swiss NSF and NCCR QSIT.

\bibliography{Normal_Transport_in_SOI_wires}

\end{document}